\documentclass[12pt]{article}
\usepackage{amsmath,amssymb,exscale}
\usepackage{psfrag,graphicx}
\input epsf

\def\gappeq{\mathrel{\rlap {\raise.5ex\hbox{$>$}}
{\lower.5ex\hbox{$\sim$}}}}
\def\lappeq{\mathrel{\rlap{\raise.5ex\hbox{$<$}}
{\lower.5ex\hbox{$\sim$}}}}
\newcommand{\Tr}{\mathop{\rm Tr}}
\def\I{\rm 1\kern-.24em l}  

\begin{document}
\topmargin -1.0cm
\oddsidemargin -0.8cm
\evensidemargin -0.8cm

\pagestyle{empty}
\begin{flushright}
UAB-FT-578\\
January 2005
\end{flushright}
\vspace*{5mm}

\begin{center}
\vspace{3.0cm}
{\Large\bf
 Chiral symmetry  breaking  from five dimensional spaces }\\
\vspace{3.0cm}
{\large Leandro  Da Rold and
Alex Pomarol}\\
\vspace{.6cm}
{\it { IFAE, Universitat Aut{\`o}noma de Barcelona,
08193 Bellaterra, Barcelona}}\\
\vspace{.4cm}
\end{center}

\vspace{1cm}
\begin{abstract}
Based on the  AdS/CFT correspondence we study    the breaking of
the chiral symmetry in QCD using   a simple five dimensional
model. The model gives  definite predictions for the spectrum of
vector mesons, their decay constants and interactions as a
function of one  parameter related to the quark condensate. We
calculate the coefficients $L_i$ of the low-energy QCD chiral
lagrangian, as well as other physical quantities for
the pions. All the predictions  are shown to be in good agreement
with the experimental data. We also show that they   are   robust
under modifications of the 5D metric in the IR, and that some 
of them arise as a  consequence of the higher-dimensional gauge
symmetry. For example, at the tree-level, we find
 $M_\rho \simeq \sqrt{3}\, g_{\rho\pi\pi} F_\pi$,
$F_\rho\simeq \sqrt{3}\,  F_\pi$
and    BR($a_1\to \pi\gamma)=0$.
\end{abstract}

\vfill
\eject
\pagestyle{empty}
\setcounter{page}{1}
\setcounter{footnote}{0}
\pagestyle{plain}
%

\section{Introduction}

The string/gauge
duality  \cite{Maldacena:1997re}
 has allowed us in the last years
to  gain  new insights into  the problem of strongly coupled gauge
theories. Although a string description of real QCD has not yet
been formulated, different string constructions have been able to
describe gauge theories with certain similarities  to QCD.
Recently, the incorporation of D7-branes in the AdS$_5\times$S$^5$
background \cite{Karch:2002sh} has allowed to address flavor
issues
\cite{Kruczenski:2003be}.

A related but  more phenomenological approach  to QCD has
consisted in extracting properties of QCD using 5D field theories
in Anti-de-Sitter (AdS)  
\cite{Polchinski:2002jw,Karch:2002xe,Boschi-Filho:2002vd,Son:2003et,Hong:2003jm}. 
This approach is based on the  AdS/CFT
correspondence \cite{Gubser:1998bc} 
that relates strongly coupled
conformal field theories (CFT) to weakly coupled 5D theories in
AdS. 
This is a more modest attempt   but, in certain regimes, it
grasps the generic features of the more involved string
constructions.

This approach   can be useful to study chiral symmetry breaking in
the vector and axial-vector sector of QCD. It is known  from the
OPE that the vector-vector current correlator for large Euclidean
momentum, $p\gg \Lambda_{\rm QCD}$, is given in the chiral limit
 by \cite{Shifman:1978bx}
\begin{equation}
\Pi_V(p^2)=p^2\left[\beta \ln\frac{\mu^2}{p^2}
+\frac{\gamma}{p^4}+\frac{\delta}{p^6}+\cdots\right]\, ,
\label{pivqcd}
\end{equation}
where $\beta\simeq  N_c/(12\pi^2)$, $\gamma\simeq\alpha_s\langle
G_{\mu\nu}^2\rangle/12\pi$ and $\delta\simeq -28\pi
\alpha_s\langle \bar qq\rangle^2/9$
 are almost momentum-independent coefficients. Similar
 expression holds for the axial-axial correlator $\Pi_A$.
Therefore  QCD behaves in Eq.~(\ref{pivqcd}) as a near-conformal
theory in the ultraviolet (UV) in which the breaking of the
conformal symmetry is given by the condensates.
The correlator $\Pi_V$, on the other hand, must have, according to
the large-$N_c$ expansion, single poles in the imaginary axis of
$p$ corresponding to  colorless vector resonances.
These properties of QCD can be implemented in a  5D theory in AdS.
The condensates $\langle{\cal O}\rangle$ are described, 
in the AdS side,  by  vacuum expectation values (VEV) of  scalars 
$\Phi$ whose  masses are related to the dimension $d$ of ${\cal O}$
by  \cite{Gubser:1998bc}  
$d=\sqrt{4+M^2_\Phi L^2}+2$ ($L$ is the AdS curvature radius), while
confinement and the  mass gap in QCD can be obtained in the AdS$_5$ by
compactifying the fifth dimension.
Alike large-$N_c$ QCD, 
 the 5D theory is also
described as a function of weakly coupled states corresponding to
the mesons. 

In this article we will present a simple 5D model to study chiral
symmetry breaking in QCD.
 We  will calculate the  vector and axial
correlators, $\Pi_V$ and  $\Pi_A$, and derive from them the masses
and decay constants of the vector, axial-vector and
pseudo-Goldstone (PGB) mesons. We will also calculate their
interactions and show some generic properties of  5D models. 
As an example, 
we will  study the electromagnetic form factor of the pions and
show how vector-meson dominance (VMD) appears. Finally, we will
derive the PGB chiral lagrangian  arising  from this 5D
model and we will give the predictions for the  $L_i$ coefficients
as well as for the PGB masses. 
We will compare all these predictions
with the experimental data.

The model presented  here can also be useful to study electroweak
symmetry breaking along the lines of Refs.~\cite{Csaki:2003zu,
Barbieri:2003pr,Luty,Agashe:2004rs}. For this reason in the
appendix we study the generic case in which
the breaking of the chiral or electroweak  symmetry arises from an
operator of dimension~$d$. This allows us  to
calculate the dependence on $d$ of the Peskin-Takeuchi
\cite{Peskin:1991sw}
  $S$ parameter.

\section{A 5D model for chiral symmetry breaking}

The 5D analog of QCD with 3 flavors consists in a theory with a
SU(3)$_L\otimes$SU(3)$_R$ gauge symmetry in the 5D bulk and a  parity defined as
the interchange $L\leftrightarrow R$. We will not consider the
extra U(1)$_{L,R}$ that involves the anomaly. The 5D metric is
defined generically as
\begin{equation}
ds^2=a^2(z)\big(\eta_{\mu\nu}dx^\mu dx^\nu-dz^2\big)\, ,
\end{equation}
where $a$ is the warp factor
that in the case of AdS$_5$ is given by
\begin{equation}
a(z)=\frac{L}{z}\, , \label{ads}
\end{equation}
where  $L$ is the AdS curvature radius. We will compactify  this
space by putting two boundaries, one  at $z=L_0$ (UV-boundary) and
another  at $z = L_1$ (IR-boundary). The theory is then defined on
the line segment $L_0 \leq z \leq L_1$ \cite{Randall:1999ee}. 
The UV-boundary
acts as a regulator necessary to obtain finite calculations. The
 limit $L_0\to  0$  should be taken after divergencies
are canceled by adding  counterterms on the UV boundary
\cite{Gubser:1998bc}. The IR-boundary is needed to introduce  a
mass gap in the theory $\sim 1/L_1$.

The only fields in the bulk that we will consider are the gauge
boson fields, $L_M$ and $R_M$,
 and a scalar field $\Phi$ transforming as a
(${\bf 3_L}$,${\bf 3_R}$) whose VEV will be  responsible for the
breaking of the chiral symmetry.
 The action is given by
\begin{equation}
S_5=\int d^4x\int dz\, {\cal L}_5\, ,
\end{equation}
where
\begin{equation}
{\cal L}_5 = \sqrt{g}\, {M_5} \Tr\left[-\frac{1}{4} {L_{MN}L^{MN}}
-\frac{1}{4}{R_{MN}R^{MN}} +\frac{1}{2}{|D_M\Phi|^2}
-\frac{1}{2}M^2_\Phi|\Phi|^2\right]\, ,
\label{l5}
\end{equation}
the covariant derivative is defined as
\begin{equation}
D_M\Phi=\partial_M \Phi+iL_M\Phi-i\Phi R_M\, ,
\end{equation}
and  $g$  is the determinant of the metric. We have defined
$L_M=L_M^aT^a$ where $M=(\mu,5)$ and  Tr$[T^aT^b]=\delta_{ab}$,
and similarly for the other fields. A coefficient $M_5$ has been
factored out in front of the lagrangian so that $1/\sqrt{M_5}$ is
the 5D expansion parameter playing the role of $1/\sqrt{N_c}$ in
QCD. We define $\Phi=S\, e^{iP/v(z)}$ where $v(z)\equiv\langle
S\rangle$ and
 $S$ corresponds to a real scalar and $P$ to a real pseudoscalar
($S\rightarrow S$ and $P\rightarrow -P$ under  $L\leftrightarrow R$).
They transform  as  ${\bf 1+8}$ under SU(3)$_V$.

Let us study $v(z)$ in the case of AdS$_5$. We  assume
$M^2_\Phi=-3/L^2$ that corresponds in the CFT  to an operator
of dimension 3 such as $\bar qq$.
 Solving the bulk equation
of motion for $S$ we get
\begin{equation}
v(z)=c_1\, z+c_2\, z^3\, ,
\label{vev}
\end{equation}
where $c_1$ and $c_2$ are two integration constants.
They  can be determined as a function
of the  value of $v(z)$  at the boundaries:
\begin{equation}
c_1=\frac{M_q L_1^3-\xi\,L_0^2}{LL_1(L_1^2-L_0^2)}\ ,\qquad
c_2=\frac{\xi-M_qL_1}{LL_1(L_1^2-L_0^2)}\, ,
\end{equation}
where we have
defined
\begin{equation}
M_q\equiv
\frac{L}{L_0}v\big|_{L_0}
\ ,\qquad \xi\equiv L\, v\big|_{L_1}\, .
\end{equation}
By the AdS/CFT correspondence, 
a nonzero $M_q$ is equivalent to  put 
 an explicit breaking of the chiral symmetry in the CFT (such
as adding quark masses). On the other hand, a nonzero value of
$\xi\propto \I$ corresponds in the chiral limit, $M_q=0$, to an
spontaneously breaking SU(3)$_L\otimes$
SU(3)$_R\rightarrow$SU(3)$_V$, playing the role of  the condensate
$\langle \bar q q\rangle$ in QCD. By substituting the  solution
Eq.~(\ref{vev}) back into the action, we obtain the vacuum energy.
Taking the limit $L_0\to 0$ while keeping $M_q$ fixed, this is
given by (up to divergent  terms)
\begin{equation}
S_4\simeq -M_5L\int d^4x\Tr\left[\frac{M^2_q}{L_1^2}
-{2}\frac{\xi M_q}{L_1^3}
+\frac{3}{2}\frac{\xi^2}{L_1^4}
+V(\xi)\frac{1}{L_1^4}
\right]\, ,
\label{energy}
\end{equation}
where we have added a potential on the IR-boundary
$V(\xi)$.
This potential is assumed to exist in order to have a nonzero $\xi$
at the minimum of  Eq.~(\ref{energy})
even in the chiral limit $M_q=0$.
Possible origins of a potential for $\xi$  are given in
 Refs.~\cite{Kruczenski:2003be}.
In the following  we will take  $\xi\rightarrow \xi\I$
where  $\xi$ will be considered   an input parameter.
Therefore the vector sector of the model has 4 parameters,
$M_5$, $M_q$, $L_1$, and  $\xi$.
As we will see $M_5$ is related to $N_c$,  $M_q$ to the quark masses
and  $1/L_1$ corresponds to the mass
gap to be related to  $\Lambda_{\rm QCD}$.
The model then has, with respect to QCD,
 only one extra parameter, $\xi$.

Few comments are in order. Using naive dimensional analysis
one can estimate that
this 5D  theory becomes strongly coupled at a scale 
$\sim 24\pi^3 M_5$. 
This implies that extra  (stringy) physics   must appear at
this scale or, equivalently,
 that  this is the 
scale that suppresses higher dimensional operators in 
Eq.~(\ref{l5}).
We estimate this scale to be around few GeV.
 Second, we are neglecting the backreaction on the metric due to the
presence of the scalar VEV. Although a nonzero  energy-momentum
tensor of  $\Phi$ will affect  the geometry of the space producing
a departure from AdS, this effect will only be relevant at  $z$
very close  to the  IR-boundary, and therefore it  will not
substantially change our results. We will comment on this in the
last section. Notice that neglecting the backreaction corresponds
to   freeze other possible condensates  that turn on in the
presence of  the quark condensate.

\section{Vector, axial-vector and PGB sectors}

We are interested in studying the vector, axial-vector and PGB
sectors. The scalar sector is more
 model-dependent and will be left  for the future.
Let us first consider the chiral limit $M_q=0$.
We take this limit  in the following way.
First we consider $c_1\rightarrow 0$ with fixed $L_0$ 
and, after performing the calculations, we take  $L_0\rightarrow 0$.
In the chiral limit we have $v\propto\I$ and then
it is convenient to 
 define  the vector and axial gauge bosons:
\begin{eqnarray}
V_M&=&\frac{1}{\sqrt{2}}\big(L_M+R_M)\, ,\nonumber\\
A_M&=&\frac{1}{\sqrt{2}}\big(L_M-R_M)\, .
\end{eqnarray}
By adding the gauge  fixing terms
\begin{equation}
\begin{aligned}
{\cal L}_{GF}^V&=-\frac{M_5
  a}{2\xi_V}\Tr\left[\partial_{\mu}V_{\mu}-\frac{\xi_V}{a}\partial_5
(a V_5)\right]^2\, ,
\\
{\cal L}_{GF}^A&=-\frac{M_5
  a}{2\xi_A}\Tr\left[\partial_{\mu}A_{\mu}-\frac{\xi_A}{a}\partial_5 (a
  A_5)-\xi_A \sqrt{2}a^2 v P\right]^2\, ,
\end{aligned}
\end{equation}
  the gauge bosons $V_{\mu}$ and
$A_{\mu}$ do not mix with the scalars $A_5$ and $P$. We will take
the limit $\xi_{V,A}\rightarrow\infty$, i.e.
\begin{equation}
P=-\frac{1}{\sqrt{2} a^3 v}\partial_5 (a\pi)\ ,\qquad   \pi\equiv A_5\, .
\label{restric}
\end{equation}
After integration by parts the 5D quadratic terms for the gauge bosons and the
pseudoscalar $\pi$ are
\begin{equation}\label{L2}
\begin{aligned}
&{\cal L}_V=\frac{aM_5}{2}\Tr\left\{V_{\mu}
\left[(\partial^2-a^{-1}\partial_5 a\partial_5) \eta_{\mu\nu} -
\partial_{\mu}\partial_{\nu}\right]V_{\nu}\right\} \, ,
\\
&{\cal L}_A=\frac{aM_5}{2}\Tr\left\{A_{\mu}
\left[(\partial^2-a^{-1}\partial_5 a\partial_5+ 2v^2a^2)
\eta_{\mu\nu} - \partial_{\mu}\partial_{\nu}\right]A_{\nu}\right\}
\, ,
\\
&{\cal L}_{\pi}=\frac{M_5}{2}\Tr\left[{a}(\partial_{\mu}\pi)^2+\frac{1
}{2 a^3 v^2}(\partial_{\mu}\partial_5 (a\pi))^2-2 v^2 a^3\left(\pi
  -\partial_5\left[\frac{1}{2a^3 v^2}\partial_5 (a\pi)\right]\right)^2\right] \, .
\end{aligned}
\end{equation}
There are also boundary terms
\begin{equation}
{\cal L}_{\rm bound}=
\frac{aM_5}{2}\Tr\left(V_{\mu}\partial_5V_{\mu}-2V_{\mu}\partial_{\mu}V_5
+ A_{\mu}\partial_5A_{\mu}
-2A_{\mu}\partial_{\mu}\pi\right)\Big|^{L_1}_{L_0} \, .
\label{bound}
\end{equation}
The IR-boundary terms can be
 canceled  by imposing  the following boundary
conditions:
\begin{equation}
\partial_5 V_{\mu}\big|_{L_1}=V_5\big|_{L_1}=
\partial_5 A_{\mu}\big|_{L_1}=\pi\big|_{L_1}=0\, .
\label{irbc}
\end{equation}
The UV-boundary conditions will be discussed later. Other
important quantities that we  will be interested are the
  vertices:
\begin{eqnarray}
{\cal L}_{VA\pi}&=&i{\sqrt{2}aM_5}
\Tr\left[A_{\mu}[\partial_5 V_{\mu},\pi]+\frac{1}{2}
A_\mu[V_\mu,A_5]\delta(z-L_0)\right]\, ,
\label{31}
\\
{\cal L}_{V\pi\pi}&=&\frac{iaM_5}{\sqrt{2}}
\Tr\left(\partial_{\mu}\pi[V_{\mu},\pi]\right)+
\frac{iM_5}{2\sqrt{2}a^3 v^2}
\Tr\left(\partial_{\mu}\partial_5(a\pi)[V_{\mu},\partial_5(a\pi)]\right)\, ,
\label{32}
\\
{\cal L}_{\pi^4}&=&\frac{M_5}{48a^9
  v^6}\Tr\left[(\partial_5(a\pi)\;\partial_{\mu}\partial_5(a\pi))^2 -
(\partial_{\mu}\partial_5(a\pi))^2 (\partial_5(a\pi))^2\right]\, .\label{33}
\end{eqnarray}

\subsection{The current-current  correlators $\Pi_{V,A}$}

In QCD the generating functional of the current-current
correlators is calculated by integrating out the quarks and gluons as
a function of the  external sources. This must be equivalent in
the large-$N_c$ limit to integrate all the colorless resonances at
tree-level. The  AdS/CFT correspondence tells us that this
generating functional is the result of  integrating out, at
tree-level, the 5D gauge fields restricted to a given UV-boundary
value:
\begin{equation}
\label{uvbc}
V_{\mu}\big|_{L_0}=v_{\mu}\ ,\qquad A_{\mu}\big|_{L_0}=a_{\mu}\, .
\end{equation}
The boundary fields $v_\mu$ and $a_\mu$ play the role of  external
sources coupled respectively to the vector and axial-vector QCD
currents. At the quadratic level the generating functional is
simple to calculate. We have to solve the equations of motion
Eqs.~(\ref{L2}) for the 5D gauge fields  restricted to the
UV-boundary condition Eq.~(\ref{uvbc}), and substitute the
solution back into the action. This leads  to 
the effective lagrangian that gives 
the generating
functional of the two-point correlators $\Pi_{V,A}$:
\begin{equation}\label{Leff}
{\cal L}_{\rm eff}
=\frac{P_{\mu\nu}}{2}\Tr\left[{v}_{\mu}\Pi_V(p^2){v}_{\nu}
+{a}_{\mu}\Pi_A(p^2) a_{\nu}\right] \, .
\end{equation}
We are working in momentum space and
$P_{\mu\nu}=\eta_{\mu\nu}-p_\mu p_\nu/p^2$.
For the AdS$_5$ space the $\Pi_V$ can be calculated analytically
\cite{Arkani-Hamed:2000ds,Agashe:2004rs}:
\begin{equation}\label{PiV}
\Pi_V(p^2)=-M_5L\frac{ip}{L_0}\frac{J_0(ip L_1)
Y_0(ip L_0)-J_0(ip L_0)Y_0(ip L_1)}
{J_0(ip L_1)Y_1(ip L_0)-J_1(ip L_0)Y_0(ip L_1)} \, ,
\end{equation}
where $J_n,Y_n$ are Bessel functions of order $n$ and $p$ is the
Euclidean momentum. For large  momentum, $pL_1\gg 1$, the
dependence on $p$ of the correlators is dictated by the conformal
symmetry and we find
\begin{equation}
\label{largep}
\Pi_V(p^2)\simeq-\frac{M_5L}{2}\, p^2 \ln(p^2L_0^2) \, .
\end{equation}
$1/L_0$ plays the role of a UV-cutoff  that can be absorbed in the
bare kinetic term of $v_\mu$. The coefficient of
Eq.~(\ref{largep}) must be matched to the QCD $\beta$-function of
Eq.~(\ref{pivqcd}). We get
\begin{equation}
M_5L=\frac{N_c}{12\pi^2}\equiv \tilde N_c\, ,
\label{m5}
\end{equation}
that   fixes the value of the 5D coupling. The next to leading
terms in the large momentum expansion of Eq.~(\ref{largep}) appear
suppressed exponentially with the momentum $\sim e^{-pL_1}$, contrary to
the QCD $\Pi_V$ correlator of Eq.~(\ref{pivqcd}). This is because
in our 5D model  the vector $V_M$ does not couple  to
$\langle\Phi\rangle^2$, and therefore  it does not feel the
breaking of the conformal symmetry coming from
$\langle\Phi\rangle^2$. In fact the only breaking of the conformal
symmetry  that  $V_M$ feels arises from the IR-boundary that
sharply cuts the AdS$_5$ space, but these effects decouple
exponentially at large momentum. To reproduce the extra terms of
Eq.~(\ref{pivqcd}), we would have to consider either higher-dimensional
operators mixing $V_M$ with $\langle\Phi\rangle^2$ or 
IR deviations from the AdS$_5$ space.

In large-$N_c$ theories the correlators $\Pi_{V,A}$
can be rewritten as a sum over narrow
resonances:
\begin{equation}\label{sigma}
\Pi_V=p^2\sum_n \frac{F_{V_n}^2}{p^2+M_{V_n}^2} \, ,\;\;\;
\Pi_A=p^2\sum_n \frac{F_{A_n}^2}{p^2+M_{A_n}^2}+{F_{\pi}^2} \, .
\end{equation}
 $F_{V_n}$ and $F_{A_n}$ are the vector and axial-vector decay
constants and the poles of $\Pi_{V,A}$ give the mass spectrum.
The correlators $\Pi_{V,A}$  calculated  via the AdS/CFT
correspondence can also be rewritten
as in Eq.~(\ref{sigma}).
For the AdS$_5$ space
the masses $M_{V_n}$ are determined by the poles of Eq.~(\ref{PiV}):
\begin{equation}\label{mVn}
J_0(M_{V_n}L_1)\simeq 0\ \longrightarrow\
M_{V_n}\simeq \left(n-\frac{1}{4}\right)\frac{\pi}{L_1} \, .
\end{equation}
For the $n=1$ resonance, the  rho meson, we have $M_\rho\simeq
2.4/L_1$ that we will use to determine the value of $L_1$
\begin{equation}
 M_\rho\simeq 770\ {\rm MeV}\ \to\ \frac{1}{L_1}\simeq 320\ {\rm MeV}\, .
\label{L1det}
\end{equation}
The  vector decay constants  are given by the residues of the poles
of $\Pi_V/p^2$. We obtain
\begin{equation}\label{FV}
F_{V_n}^2=
\tilde  N_c\frac{\pi M_{V_n}}{L_1}
 \frac{Y_0(M_{V_n} L_1)}{J_1(M_{V_n}L_1)} \, .
\end{equation}
Using Eqs.~(\ref{m5}), (\ref{L1det}) and (\ref{FV})  we obtain
$F_{V_1} \simeq  140$ MeV to be compared with the experimental
value $F_\rho=153$ MeV.
For the higher resonances we obtain $F_{V_{2,3}} \simeq  210, 270$ MeV.

\begin{figure}[htbp]
  \centering
  \psfrag{mA1}{$m_{A_1}$[GeV]}
  \psfrag{x}{$\xi$}
  \includegraphics[width=8cm]{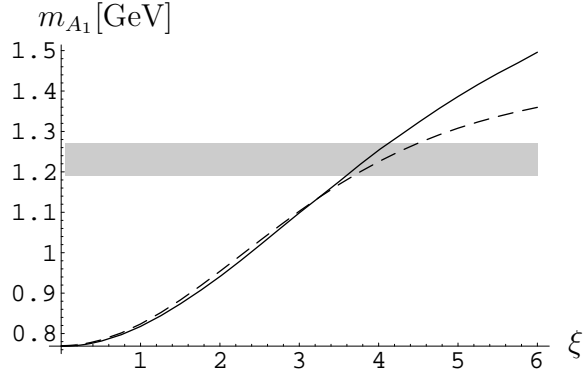}
  \caption{\textit{Mass of the first axial-vector resonance
as a function of $\xi$.
The solid line is the exact result, while the
  dashed line corresponds to  the approximate value coming from
Eq.~(\ref{massa1}).
  The shadow band  shows  the experimental value $M_{a_1}=1230\pm 40$ MeV.}}
  \label{massa1plot}
\end{figure}

The correlator $\Pi_{A}$ depends on the $z$-dependent mass of $A_\mu$
 and cannot be calculated
analytically. Numerical analysis is therefore needed to  obtain
the masses and decay constants of the axial-vector mesons.
Analytical formulas, however, can be obtained if we approximate
the  5D  mass of $A_\mu$ as a IR-boundary 4D mass, $M_{\rm IR}$.
This is expected to be  a good approximation since the scalar VEV
$v(z)$ that gives a mass to $A_\mu$  grows towards the IR-boundary
as $v(z)\simeq (z/L_1)^3\xi/L$ and is only relevant for values of
$z$ close to the IR-boundary. The value of $M_{\rm IR}$ is
determined by
\begin{equation}
\int^{L_1}_{L_0}
dz\, a^3(z) M^2_{\rm IR} A_\mu\,  \delta(z-L_1)=
M_5\int^{L_1}_{L_0}
dz\, 2a^3(z)v^2(z) A_\mu\, .
\label{mbound}
\end{equation}
The effect of a IR-boundary mass is simply to change the
IR-boundary condition from Eq.~(\ref{irbc}) to
$\big[M_5\partial_5+a^2M^2_{\rm IR}\big] A_{\mu}\big|_{L_1}=0$
\cite{Gherghetta:2000qt}, and therefore the equation that
determines the mass spectrum changes from Eq.~(\ref{mVn}) to
\begin{equation}
J_0(M_{A_n}L_1)\simeq -\int^{L_1}_{L_0}
dz\, \frac{2a^3(z)v^2(z)}{M_{A_n}L}z J_1(M_{A_n}z)\, .
\label{massa1}
\end{equation}
In Fig.~\ref{massa1plot} we  show the value of the mass of the
lowest state as a function of $\xi$. We compare  the exact
numerical value of $M_{A_1}$  and the approximate value coming
from  Eq.~(\ref{massa1}). We see that the difference is below the
$10\%$. For  $\xi\simeq 4$   we find that
 $M_{A_1}$ coincides with the experimental mass of the  $a_1$, $M_{a_1}\simeq 1230$ MeV.
We then see that the experimental data favor values of $\xi$
around 4. For this value $\xi\simeq 4$ we also find that
$F_{A_1}\simeq 160$ MeV. For the second resonance we find, for
$\xi=4$, $M_{A_2}\simeq 2$ GeV and $F_{A_2}\simeq 200$ MeV. For
heavier axial-vector resonances the right-hand side of
Eq.~(\ref{massa1}) can be neglected and then their masses approach
to the values of the
 vector masses Eq.~(\ref{mVn}) (and similarly for the decay
constants).

In the large  and small momentum limits the correlator $\Pi_A$
can  also be calculated analytically without the need of
the above approximation.
Furthermore these analytical expressions simplify enormously
 if $\xi\gg 1$.
In this limit we find that  the dependence of $\Pi_A$ on $\xi$  is
simply dictated by the conformal symmetry. For small $p$, we have
\begin{equation}
\Pi_A(p^2)=\Pi_A(0)+p^2 \Pi_A^{\prime}(0)+{\cal O}(p^4)\, ,
\label{AlowE}
\end{equation}
where  for $\xi\gg 1$
\begin{eqnarray}
\Pi_A(0)&=&F_\pi^2\simeq
 \frac{2^{5/3}\pi}
{3^{1/6}\Gamma(\frac{1}{3})^2}\frac{\tilde N_c\, \xi^{2/3}}{L_1^2}
\, ,
\label{fpi}\\
\Pi_A^{\prime}(0)&\simeq&
-\tilde N_c\left[
\ln\frac{L_0}{L_1}
+\ln \xi^{1/3}
+
\frac{4\gamma+\pi\sqrt{3}-\ln 12}{12}
\right]
\, .
\label{PiAp}
\end{eqnarray}
From Eq.~(\ref{fpi}), making use of   Eqs.~(\ref{m5}) and
(\ref{L1det}), we get
\begin{equation}
F_\pi\simeq 87 \left(\frac{\xi}{4}\right)^{\frac{1}{3}}\ {\rm MeV}\, ,
\label{fpinum}
\end{equation}
in excellent agreement with the experimental value for $\xi\simeq
4$. We have checked  that the approximate value of $F_\pi$ 
Eq.~(\ref{fpinum}) differs from the exact value  by less than
10$\%$ if $\xi\gappeq 3$. Adding an  explicit breaking of the
chiral symmetry, $M_q\not=0$, gives an extra contribution to
$\Pi_A(0)$. By expanding  around $M_q=0$, we obtain
  $\Pi_A(0)=\Pi_A^{(0)}(0)+{M_q}{L_1}\Pi^{(1)}_A(0)+\cdots$, where,
in the limit $\xi\gg 1$,  $\Pi_A^{(0)}$ is given by Eq.~(\ref{fpi}) and
\begin{equation}
\Pi^{(1)}_A(0)\simeq \frac{
2^{8/3} 3^{2/3}\pi}
{\Gamma(\frac{1}{6})^2}\frac{\tilde N_c\, \xi^{1/3}}{L_1^2}
\left(1-\frac{2\Gamma(\frac{1}{6})}{6^{4/3}
\sqrt{\pi}} \frac{1}{\xi^{2/3}}
\right)
\, .
\label{PiA1}
\end{equation}
Eq.~(\ref{PiA1}) gives a contribution to $F_\pi$ proportional to
the quark masses $M_q$.

In the large momentum limit
    $\Pi_{A}$ is given in the chiral limit by 
\begin{equation}
\Pi_A(p^2)=-p^2\left[
\frac{\tilde N_c}{2} \ln(p^2L_0^2)+\frac{c_6}{p^6}
+{\cal O}(\frac{1}{p^{12}})\right]\ ,
\qquad {\rm where}\ \ \
c_6=-\frac{16}{5}\frac{\tilde N_c\, \xi^2}{L_1^6}
\, .
\label{PiAlarge}
\end{equation}
As we said before,  corrections  to Eq.~(\ref{PiAlarge}) are
expected 
if the  5D  metric departs in the IR from AdS.
Nevertheless,
these corrections 
cancel out in 
the left-right correlator $\Pi_{LR}=\Pi_V-\Pi_A$.
Therefore, at large momentum 
we have
\begin{equation}
\label{pilr}
\Pi_{LR}(p^2) \simeq  \frac{c_6}{p^4}+\cdots\, ,
\end{equation}
independently of variations in the AdS$_5$  metric.
Eq.~(\ref{PiAlarge}) gives
\begin{equation}
 c_6\simeq
-1.4\times 10^{-3}\left(\frac{\xi}{4}\right)^2 {\rm GeV}^6\, ,
\end{equation}
 to be compared with
the QCD  value $c_6=-4\pi\alpha_s \langle\bar{q}q\rangle^2\simeq
-1.3\times 10^{-3}$ GeV$^6$ extracted from
 the evaluation of the condensate of Ref.~\cite{Jamin:2002ev}.
We must notice, however, that Eq.~(\ref{pilr}) will be affected by
  higher-dimensional operators such as 
Tr$[\Phi^\dagger L_{MN}\Phi R^{MN}]$.

\subsection{Vector meson interactions}
\label{vmin}

\begin{figure}[htbp]
  \centering
  \psfrag{V1}{\small{V$_1$}}
  \psfrag{V2}{\small{V$_2$}}
  \psfrag{A1}{\small{A$_1$}}
  \psfrag{Pi}{\small{$\pi$}}
  \psfrag{x}{\Large{$\frac{z}{L_1}$}}
  \includegraphics[width=8cm]{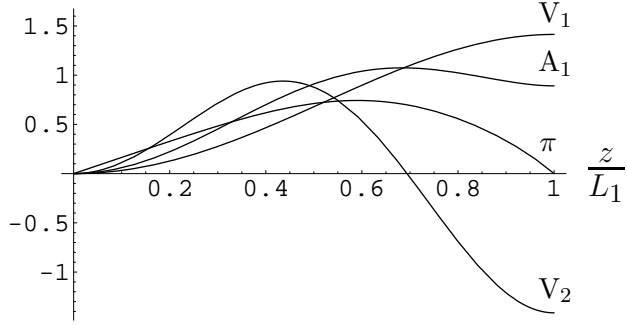}
  \caption{\textit{Wave-function  of  the $n=1,2$ vector
  resonance, the $n=1$  axial-vector resonance and the PGB.}}
  \label{wfplot}
\end{figure}

To calculate the couplings between the resonances
it is convenient to perform a
 Kaluza-Klein (KK) decomposition
of  the 5D fields. This consists in expanding the fields in
a tower of 4D mass-eigenstates,
$V_\mu(x,z)=
 \frac{1}{\sqrt{M_5L}}\sum_n
f^V_{n}(z)V_\mu^{(n)}(x)$,
and equivalently for the other fields.
To cancel
the  UV-boundary terms of Eq.~(\ref{bound}) we impose
\begin{equation}
V_{\mu}\big|_{L_0}=
 A_{\mu}\big|_{L_0}=0\, .
\label{UVbc}
\end{equation}
For the electromagnetic subgroup  of SU(3)$_V$, however, we must consider
the boundary condition $\partial_5 V_{\mu}\big|_{L_0}=0$ in order
to have   a  massless mode in the spectrum, the photon,  whose
wave-function satisfies
\begin{equation}
\partial_5 f_0^{V}=0\, .
\label{foton}
\end{equation}
In  the limit $L_0\rightarrow 0$ this state becomes
non-normalizable since its kinetic term diverges as $\tilde N_c\ln{(L_1/L_0)}$.
To  keep it as a dynamical field, we can fix
$1/L_0$    to a large but finite value. 
In the absence of UV-boundary  kinetic terms, this value of $1/L_0$ 
defines    the scale 
of the  Landau pole \cite{Arkani-Hamed:2000ds}:
\footnote{This is
equivalent  to add a kinetic term on the UV-boundary with coupling
$1/e_0^2=\tilde N_c\ln{(L_0\mu)}+1/e^2(\mu)$ 
that  cancels, in the limit $L_0\rightarrow 0$,  
 the
divergence in the kinetic term of the massless mode  
and normalizes this state.}
\begin{equation}
\frac{1}{e^{2}(\mu)}=-\tilde N_c\ln{(L_0\mu)}\, .
\label{e2}
\end{equation}
The wave-functions of the   KK modes $V^{(n)}_\mu$ $(n\not=0)$ are
given for the AdS$_5$  case by
\cite{Gherghetta:2000qt}:
\begin{equation}
f_n^{V}(z)=\frac{z}{N_{V_n}L_1}\left[ J_1(M_{V_n}z)+b(M_{V_n})
Y_1(M_{V_n}z)\right] \ \ \ \stackrel{L_0\to 0} {\longrightarrow }\ \ \
 \frac{z}{N_{V_n}L_1}
J_1(M_{V_n}z)
\, , \label{wfv}
\end{equation}
where $b(M_{V_n})=-\frac{J_{1}(M_{V_n}L_0)}{Y_{1}(M_{V_n}L_0)}$ and
$N_{V_n}$ is a constant fixed by canonically normalizing the field.
The masses $M_{V_n}$ are determined by the condition $\partial_5
f_n^{V}(z)\big|_{L_1}=0$ that coincides with the poles of
Eq.~(\ref{PiV}). For the vector KK modes associated to the
electromagnetic subgroup, we have
$b(M_{V_n})=-\frac{J_{0}(M_{V_n}L_0)}{Y_{0}(M_{V_n}L_0)}$. In this
case the KK masses  are different by factors of order $e^2$ from
the values of  Eq.~(\ref{mVn}).
 This is expected since the KK
modes are mass-eigenstates and their masses  incorporate
corrections
 due to the  mixing of the resonances in Eq.~(\ref{sigma})
with the photon.
In Fig.~\ref{wfplot} we plot the wave-function of the first two KK modes.

For the axial-vector  $A_\mu$ there is no massless mode.
The KK wave-functions
 cannot be obtained analytically and one must rely
in numerical analysis.
In Fig.~\ref{wfplot} we plot the
wave-function of the first KK mode, the $a_1$, for the
 AdS$_5$ case. Throughout this section 
we will work in the chiral limit.

Finally,  the pseudoscalar fields $\pi$       have also a 4D
massless mode corresponding to the PGB arising from  chiral
symmetry breaking. Their  wave-functions are determined by
\begin{equation}
\label{pion}
\left[\frac{1}{a}
-\partial_5\left(\frac{1}{2a^3v^2}\partial_5 \right)\right]af_0^{\pi}=0 \, .
\end{equation}
For
AdS$_5$ we obtain
\begin{equation}
f^{\pi}_0=\frac{z^3}{L_1^3 N_0}\left[I_{2/3}\left(\frac{\sqrt{2}\xi}{3}
\frac{z^3}{L_1^3}
\right)-
\frac{I_{2/3}\left(\sqrt{2}\xi/3\right)}{K_{2/3}\left(\sqrt{2}\xi/3\right)}
K_{2/3}\left(
\frac{\sqrt{2}\xi}{3}
\frac{z^3}{L_1^3}
\right)\right] \, ,
\end{equation}
where the constant $N_0$ canonically normalizes the field (this is
fixed by $\frac{1}{2a^2v^2L}f^{\pi}_0\partial_5
(af^{\pi}_0)|_{L_0}=1$). Here an after we will denote by $\pi$
only the massless modes, the PGB. Their wave-functions are shown
in Fig.~\ref{wfplot}.

The interactions between the different resonances are easily
obtained from  Eqs.~(\ref{31})-(\ref{33}) and  integrating over
$z$ with the corresponding  wave-functions.
Here we present some
phenomenologically relevant examples.
 The first one is the
coupling of the photon to $A^{(n)}_\mu\pi$. 
Using  Eqs.~(\ref{31}), (\ref{UVbc})  and
(\ref{foton}), we obtain that this coupling is zero:
\begin{equation}
  \label{eq:afp}
g_{A_n\gamma\pi}=0\, .
\end{equation}
Eq.~(\ref{eq:afp}) is a consequence of 
 electromagnetic gauge invariance which implies that 
$p^\nu {\cal M}_{\mu\nu}=0$ where $p^\nu$ is the momentum of the photon
and ${\cal M}_{\mu\nu}$ is the vertex $A^{(n)}_\mu\gamma_\nu \pi$.
In the 5D model  of
Eq.~(\ref{l5}), in which only dimension 4 operators are considered,
we have at tree level that 
${\cal M}_{\mu\nu}$ can only be proportional to $g_{\mu\nu}$
and then Eq.~(\ref{eq:afp}) follows from the  
condition of gauge invariance.
Eq.~(\ref{eq:afp}) 
has the interesting consequence that,
at the leading order in large-$N_c$,
the branching ratio of  $a_1\to\gamma\pi$ vanishes.
This coupling, however,  could be induced from 
5D higher-dimensional operators or quantum loop effects.

\begin{figure}[htbp]
  \centering
  \psfrag{g1aprox}{\small{$g_{1\pi\pi}^{app}$}}
  \psfrag{g1}{\small{$g_{1\pi\pi}$}}
  \psfrag{g2}{\small{$g_{2\pi\pi}$}}
  \psfrag{g3}{\small{$g_{3\pi\pi}$}}
  \psfrag{x}{$\xi$}
  \includegraphics[width=8cm]{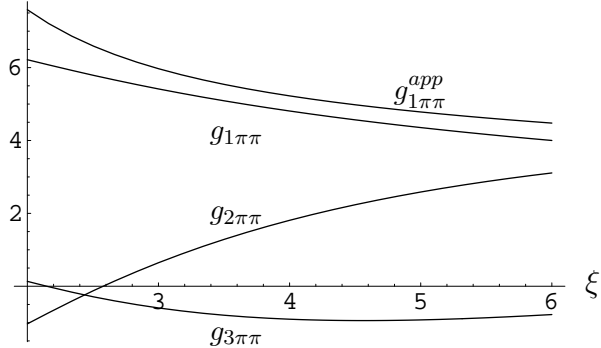}
  \caption{\textit{Coupling of the $n=1,2,3$ vector resonance to two
  PGB. We also show the approximate value for $n=1$
given by
  $g_{1\pi\pi}^{app}=M_{V_1}/(\sqrt{3}\,F_{\pi})$
 -see  Eq.~(\ref{KSRF2}).}}
  \label{rppplot}
\end{figure}

Another example is the vector coupling to two PGB. From
Eq.~(\ref{32})  we get
\begin{equation}
{\cal L}_{V_n\pi\pi}=i \frac{g_{n\pi\pi}}{\sqrt{2}}
\Tr\big(\partial_{\mu}\pi[V^{(n)}_{\mu},\pi]\big)\, ,
\end{equation}
where $g_{n\pi\pi}$ is given by
\begin{equation}
g_{n\pi\pi}=\frac{ 1}{\sqrt{M_5L^3}}\int dz \,
af_n^V\left[(f_0^{\pi})^2+ \frac{(\partial_5
(af_0^{\pi}))^2}{2a^4v^2}\right] \, . \label{rpp}
\end{equation}
In Fig.~\ref{rppplot} we show the coupling of the first three KK
modes as a function of $\xi$ for the AdS$_5$ case. One can see
that the heavier is the KK mode (larger $n$), the smaller is its
coupling to PGB. This can be understood  as a consequence   of the
 increase in the oscillations of the KK wave-function as $n$
increases (see Fig.~\ref{wfplot}), that implies
 a smaller contribution to  the integral Eq.~(\ref{rpp}) for larger $n$.

Finally, we consider the four-pion interaction. It receives
contributions coming from the exchange of vector resonances,
scalar resonances, and the  four-interaction of Eq.~(\ref{33}).
At the  order $p^2$, the chiral symmetry
tells us that this coupling must be
$(1/12F^2_\pi)\Tr[(\pi\overleftrightarrow{\partial_\mu}\pi)^2]$. This implies
the following sum rule:
\begin{equation}\label{wsr1}
{g}_{\pi^4}+\sum_n
\frac{g_{n\pi\pi}^2}{M_{V_n}^2}=\frac{1}{3F_{\pi}^2} \, ,
\end{equation}
where  ${g}_{\pi^4}$  denotes   the scalar contributions and the
four-interaction. We find that for values
$\xi\gappeq 3$ the contribution
${g}_{\pi^4}$ is small and only the vector contribution dominates.
This is saturated (at the   $90\%$ level) by the first resonance,
the rho meson, leading to the  following approximate relation
\begin{equation}
\label{KSRF2}
M_{\rho}^2 \simeq 3 F_{\pi}^2 g_{\rho\pi\pi}^2\, .
\end{equation}
In Fig.~\ref{rppplot} we plot  the approximate  value of
$g_{1\pi\pi}$  that arises from Eq.~(\ref{KSRF2}), and it is shown
that the difference from its exact value is only  $\sim 10\%$.
Eq.~(\ref{KSRF2}) differs by a factor $2/3$ from   the KSRF
relation \cite{Kawarabayashi:1966kd}, $M_{\rho}^2 \simeq 2
F_{\pi}^2 g_{\rho\pi\pi}^2$, that is known to be experimentally
very successful. The approximate relation Eq.~(\ref{KSRF2}) had
been found previously in a specific  extra dimensional model
\cite{Son:2003et}.   We have shown here, however,  that it is a
general  prediction  of 5D models independent of the space
geometry. It only relies on the 5D gauge symmetry that forbids
terms with four $A_5$.

\subsection{The electromagnetic form factor of the pion}

\begin{figure}[htbp]
  \centering
  \psfrag{Fpip}{${\cal F}_{\pi}(p)$}
  \psfrag{pGeV}{{$p\ [{\rm GeV}]$}}
  \includegraphics[width=8cm]{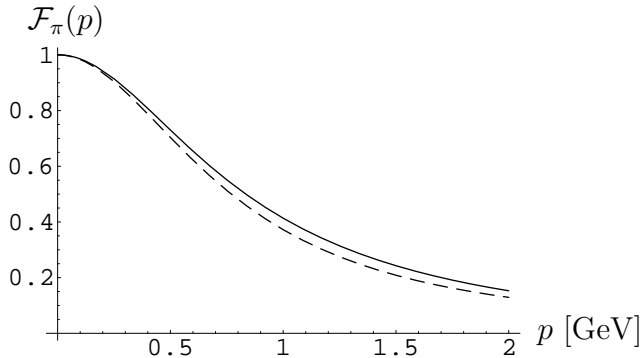}
  \caption{\textit{Electromagnetic pion form factor
as a function of the transfer momentum $p$ for $\xi=4$.
The solid line is the exact result, while the
  dashed line is obtained by considering only the rho meson (VMD).}}
  \label{formfactorplot}
\end{figure}

The electromagnetic form factor of the pion,  ${\cal F}_\pi(p)$
where $p$ is the  momentum transfer, corresponds to the coupling
of the pion to the external vector field $v_\mu$. In the 5D
picture   the pion does not couple directly to $v_\mu$ but only
through the  interchange of the vector resonances. This is because
the pion wave-function is zero at  the UV-boundary  and therefore
the pion can only couple to the UV-boundary fields through the KK
states. This implies that the form factor of the pion  can be
written as
\begin{equation}
{\cal F}_{\pi}(p)
=\sum_n g_{n\pi\pi}\frac{M_{V_n}F_{V_n}}{p^2+M_{V_n}^2} \, .
\label{wsr2}
\end{equation}
The quantization of the electric charge of the pion implies   
${\cal F}_\pi(0)=1$
from which one can derive
the sum rule $\sum_n  g_{n\pi\pi} F_{V_n}/M_{V_n}=1$.
Above we saw that the coupling $g_{n\pi\pi}$
and the inverse of the mass decrease as $n$
increases implying that this sum rule is mostly dominated by the first resonance
and therefore
\begin{equation}
  g_{\rho\pi\pi}F_{\rho}\simeq M_{\rho} \, .
\label{KSRF1}
\end{equation}
For $\xi\simeq 4$, this relation is fulfilled at the $88\%$ level.
For larger values of $\xi$, however, Eq.~(\ref{KSRF1}) is
not so well satisfied since the contribution of the second resonance becomes
important.
Eq.~(\ref{KSRF1}) together with Eq.~(\ref{KSRF2}) allows us to write
a relation
between the $\rho$ and $\pi$  decay constants
\begin{equation}
F_\rho\simeq \sqrt{3}F_\pi\, .
\end{equation}
At large momentum 
the contribution of  each  $n>1$
 resonance
 to
${\cal F}_{\pi}(p)$ becomes sizable 
since the
small value of $g_{n\pi\pi}$ for $n>1$ is compensated by the large
value of $M_{V_n}F_{V_n}$. Nevertheless, the total contribution  coming
from summing over all the  modes 
with $n>1$ approximately cancels out, implying that
the  rho  meson dominates  in Eq.~(\ref{wsr2}) even at large
momentum. This can be seen in Fig.~\ref{formfactorplot} where we
compare the exact result for ${\cal F}_\pi (p)$ to the result in which
only  one resonance is considered
${\cal F}_\pi(p)=M^2_\rho/(p^2+M^2_\rho)$. 
The cancellation  of the contribution of the heavy
modes to ${\cal F}_\pi(p)$  is a consequence of the conformal
symmetry.
At large momentum transfer the conformal symmetry 
 tells us that the electromagnetic form factor of
a scalar  hadron drops as $1/p^{(2\tau_h-2)}$ where 
$\tau_h=$Dim$[{\cal O}_h]-s$ 
being ${\cal O}_h$   the local operator that creates the hadron
from the vacuum and $s$  the  spin of the operator \cite{Polchinski:2002jw}. 
For the case of the pion
we have that $\tau_h=2$ 
(where ${\cal O}_h$ is  the axial-vector current operator)
and then ${\cal F}_{\pi}(p)$ must drop as
$1/p^2$.
This large momentum behaviour coincides with that
of the rho contribution to ${\cal F}_{\pi}(p)$.

The hypothesis that the electromagnetic form factor of the pion is
dominated by  the rho meson, that goes under name  of VMD,  was
proposed long ago. It did not have any theoretical motivation, but
it led to a  good agreement with experiments tough. We have seen,
however,  that VMD  in ${\cal F}_\pi(p)$ appears as an unavoidable
consequence of  this 5D model for $\xi\sim 4$ 
(see also Ref.~\cite{Son:2003et}).

\subsection{The effective lagrangian for the $\rho$ meson}

We have seen that the rho meson gives the largest contribution to
the pion couplings.
Therefore in order to obtain the chiral lagrangian for the PGB,
it is  convenient to write
first the effective lagrangian
for the rho meson.

In order to make contact with the literature
\cite{Ecker:1989yg},
 we will write
the effective lagrangian  not in the mass-eigenstate basis but
in the basis defined by
 $v_\mu$ as in Eq.~(\ref{uvbc}) and the rho field $V_\mu$
transforming under the SU(3)$_V$ symmetry as $V_\mu\to hV_\mu
h^\dagger+i/g\, h\partial_\mu h^\dagger$. From now on we will
follow the notation and definitions of Ref.~\cite{Ecker:1989yg}.
The effective lagrangian for $V_\mu$ invariant under the chiral
symmetry can be written  as
\begin{equation}
{\cal L}_V
= -\frac{1}{4}\Tr[V_{\mu\nu}V^{\mu\nu}]
+\frac{1}{2}M^2_\rho\Tr\big[V_\mu-\frac{i}{g}
\Gamma_\mu\big]^2-
\frac{\widetilde F_\rho}{2\sqrt{2}M_\rho}
\Tr[V_{\mu\nu}f^{\mu\nu}_+]+\cdots \, ,
\label{lrho}
\end{equation}
where
\begin{equation}
\Gamma_\mu =
\frac{1}{2} \left\{ u^\dagger (\partial_\mu - iR_\mu) u +
  u(\partial_\mu - iL_\mu) u^\dagger \right\}\ ,\ \ \ \ \
f^{\mu\nu}_+ = u F_L^{\mu\nu} u^\dagger +
u^\dagger   F_R^{\mu\nu} u \, ,
\label{connections}
\end{equation}
and  $u^2=U$ being  $U$ a parametrization of the PGB:
\begin{equation}\label{U}
U=e^{i\sqrt{2}\, {\pi}/{F_\pi}}\, .
\end{equation}
The lagrangian Eq.~(\ref{lrho}) does not contain
all possible chiral terms of ${\cal O}(p^4)$.
We have neglected couplings between
$\pi$ and $V_\mu$ involving more than one derivative
since these couplings do not arise from a 5D lagrangian.
(We have also neglected trilinear couplings between vectors
since they do not play any role in our analysis).

Matching the above  lagrangian with the 5D AdS theory we obtain
\begin{eqnarray}
\frac{1}{g}&=&2\sqrt{2}g_{\rho\pi\pi}\frac{F^2_\pi}{M^2_\rho}\, ,\nonumber\\
\widetilde F_\rho&=&F_\rho-\frac{M_\rho}{\sqrt{2}g}\, .
\label{maa}
\end{eqnarray}
Notice that our 5D  model predicts 
 a nonzero value for $\widetilde F_\rho$ 
(we find $\widetilde F_\rho \simeq 40$ MeV for $\xi=4$)
differently  from 
Ref.~\cite{Ecker:1989yg}
or models where the rho is considered a Yang-Mills field
\cite{Bando:1984ej} in which  one has $\widetilde F_\rho=0$.

\section{The  chiral lagrangian for the PGB}

By integrating all the heavy  resonances  we can obtain the effective
lagrangian for the PGB.
This lagrangian is fixed by the chiral symmetry up to some
unknown coefficients.
In these section we will give the prediction of the AdS$_5$
model for  these coefficients.

The chiral lagrangian
for the  PGB $\pi$ can be written as a function of $U$ defined
in  Eq.~(\ref{U}).  Up to  ${\cal O} (p^2)$, this is given by
\begin{equation}
{\cal L}_2
=\frac{F_\pi^2}{4}\,
\Tr\big[ D_\mu U^\dagger D^\mu U \, + \, U^\dagger\chi  \,
+  \,\chi^\dagger U\big]\, ,
\label{l2}
\end{equation}
where
\begin{equation}
D_\mu U = \partial_\mu U - i R_\mu U + i U L_\mu\, ,
\end{equation}
and
\begin{equation}
\chi  = \, 2 B_0 \, (M_q + i p)\ ,\qquad M_q={\rm Diag}(m_u,m_d,m_s)\, .
\label{chi}
\end{equation}
The constant $B_0$ is related to the quark condensate by $\langle
\bar q q\rangle=-B_0F_\pi^2$.
In the AdS$_5$ model $F_\pi$  is given by
Eq.~(\ref{fpi}),
while matching Eq.~(\ref{l2})
in the unitary gauge ($U=1$) to Eq.~(\ref{energy}) we get
\begin{equation}
F^2_\pi B_0=\frac{2\tilde N_c\, \xi}{ L_1^3}\, ,
\label{B0}
\end{equation}
that determines the PGB masses:
\begin{equation}
(m^2_\pi)_{ab}=2B_0\Tr\left[M_q T_aT_b\right]\, .
\end{equation}
From the pion mass $m_{\pi^0}\simeq 135$ MeV, we obtain $m_u+m_d\simeq 20.5$ MeV
for $\xi=4$.

At the    ${\cal O}(p^4)$ the
chiral lagrangian has extra terms  given by \cite{Gasser:1984gg}
\begin{eqnarray}
{\cal L}_4 &=&
L_1 \,{\rm Tr}^2\big[ D_\mu U^\dagger D^\mu U\big] +
L_2 \,\Tr\big[ D_\mu U^\dagger D_\nu U\big]
   \Tr\big[ D^\mu U^\dagger D^\nu U\big]
+ L_3 \,\Tr\big[ D_\mu U^\dagger D^\mu U D_\nu U^\dagger
D^\nu U\big]\,
\nonumber\\
&+&  L_4 \,\Tr\big[ D_\mu U^\dagger D^\mu U\big]
   \Tr\big[ U^\dagger\chi +  \chi^\dagger U \big]
+ L_5 \,\Tr\big[ D_\mu U^\dagger D^\mu U \left( U^\dagger\chi +
\chi^\dagger U \right)\big]\,
\nonumber\\
&+& L_6 \,{\rm Tr}^2\big[ U^\dagger\chi +  \chi^\dagger U \big]
+ L_7 \,{\rm Tr}^2\big[ U^\dagger\chi -  \chi^\dagger U \big]
+  L_8 \,\Tr\big[\chi^\dagger U \chi^\dagger U
+ U^\dagger\chi U^\dagger\chi\big]
\nonumber \\
&-& i L_9 \,\Tr\big[ F_R^{\mu\nu} D_\mu U D_\nu U^\dagger +
     F_L^{\mu\nu} D_\mu U^\dagger D_\nu U\big]
+  L_{10} \,\Tr\big[ U^\dagger F_R^{\mu\nu} U F_{L\mu\nu} \big]
\, .
\label{l4}
\end{eqnarray}
The coefficients $L_{1,2,3}$ are responsible for  four-pion
interactions  at ${\cal O}(p^4)$, while $L_9$ gives a contribution
to the electromagnetic form factor of the pion at ${\cal O}(p^2)$.
From the discussion of the previous section we know  that the
dominant contribution to these processes arises from  the  rho
meson exchange. Therefore the main contribution to $L_{1,2,3,9}$
will arise by
 integrating out this particle.
Using the effective lagrangian Eq.~(\ref{lrho}) with
Eqs.~(\ref{maa}), we obtain \footnote{ These coefficients are
induced after performing the redefinition
 $V_\mu\rightarrow V_\mu+i\Gamma_\mu/g$
in Eq.~(\ref{lrho}).
After this redefinition
the rho meson couples to the pion only  at ${\cal O}(p^3)$ and then
it does not  induce a contribution to Eq.~(\ref{l4}) when
it is  integrated  out.}
\begin{eqnarray}
L_1&=&\frac{g^2_{\rho\pi\pi} F^4_\pi}{8M^4_\rho}\, ,
\\
L_2&=&2L_1\ ,\qquad L_3=-6L_1\, ,\\
L_9&=&\frac{ g_{\rho\pi\pi}F_\rho F^2_\pi}{2M^3_\rho}\, .
\end{eqnarray}
Using Eqs.~(\ref{KSRF2}) and (\ref{KSRF1}), we get
\begin{equation}
L_1\simeq\frac{F^2_\pi}{24M^2_\rho}\ ,\qquad
L_9\simeq\frac{F^2_\pi}{2M^2_\rho}\, .
\end{equation}
The coefficients $L_{4,6}$ are zero at the tree-level (leading
order in the large-$N_c$ expansion), while
 the calculation of  $L_{7,8}$ will be left for the future.
$L_7$ involves the U(1) anomaly and $L_8$ only receives
contributions from the scalar sector. $L_5$ and $L_{10}$  can be
calculated from the correlators $\Pi_{V,A}$:
\begin{eqnarray}
L_{5}&=&\frac{L_1}{8B_0}\Pi_A^{(1)}(0)\, ,\ \ \ \ \ ({\rm only\ if}\
L_4=0)\, ,\label{l5def}\\
L_{10}&=&\frac{1}{4}\big[\Pi^\prime_{A}(0)-\Pi^\prime_{V}(0)\big]\, ,
\end{eqnarray}
where $L_1$  in Eq.~(\ref{l5def}) is the one defined in Eq.~(\ref{L1det}).
From Eqs.~(\ref{PiV}), (\ref{fpi})-(\ref{PiA1}) and (\ref{B0})
we obtain for $\xi\gg 1$ 
\begin{eqnarray}
L_5&\simeq&\tilde N_c
\frac{2\pi^3}{\sqrt{3}
\Gamma(\frac{1}{3})^6}\left[1-\frac{2\Gamma(\frac{1}{6})}{6^{4/3}
\sqrt{\pi}}\frac{1}{\xi^{2/3}}
\right]
\simeq 2.5\cdot 10^{-3}\left[1-0.23\left(\frac{4}{\xi}\right)^{\frac{2}{3}} \right]
\, ,
\\\nonumber
\\
L_{10}&\simeq&-\frac{\tilde N_c}{4}\left[
\ln \xi^{1/3}
+
\frac{4\gamma+\pi\sqrt{3}-\ln12}{12}
\right]
\simeq -5.7\cdot 10^{-3}
\left[\ln\left(\frac{\xi}{4}\right)^{\frac{1}{3}}+1\right]
\, .
\end{eqnarray}
In Table~1 we compare the experimental values of $L_i$ with the
predictions  of our AdS$_5$ model for the value $\xi=4$. We give
the exact values of our  predictions although we find that the
predictions in the limit $\xi\gg 1$ differ  by less than a $10\%$
from the exact results. Comparing the predictions with the
experimental values we find that the discrepancy is always below
the $30\%$.

\begin{table}[htb]
\begin{center}
\begin{tabular}{cccc}
\hline
  & {\rm Experiment}
    &{\rm AdS$_5$}
\\ \hline
 $L_1$ & $0.4\pm0.3$ &
  $0.4$
\\
 $L_2$ & $1.4\pm0.3$ &
       $0.9$
\\
 $L_3$ & $-3.5\pm1.1$ &
        $-2.6$
\\
 $L_4$ & $-0.3\pm0.5$ &
 $0.0$
\\
 $L_5$ & $1.4\pm0.5$ &
            $1.7$
\\
 $L_6$ & $-0.2\pm0.3$ &
 $0.0$
\\
 $L_9$ & $6.9\pm0.7$ &
      $5.4$
\\
 $L_{10}$ & $-5.5\pm0.7$
& $-5.5$
\\
\hline
\end{tabular}
\caption{\it Experimental values of the $L_i$
at the scale $M_\rho$
in units of $10^{-3}$ \cite{Pich:1998xt} and
the predictions of the AdS$_5$ model  for the value $\xi= 4$.}
\end{center}
\end{table}

Finally, we  also calculate
the coefficient of the
operator $\Tr[Q_RUQ_LU^\dagger]$
responsible for
the electromagnetic pion mass difference
($Q_{L,R}$ are the left- and right-handed
charges) \cite{Ecker:1988te}.
This coefficient is given by
$e^2C=(m^2_{\pi^+}-m^2_{\pi^0})F^2_\pi/2$ where
\begin{equation}\label{mPi}
m_{\pi^+}-m_{\pi^0}\simeq \frac{3 \alpha}{8 \pi m_{\pi}
  F_{\pi}^2}\int_0^{\infty} dp^2 \big(\Pi_A-\Pi_V\big) \, .
\end{equation}
Taking $\Pi_V$ from  Eq.~(\ref{PiV}) and calculating  $\Pi_A$
numerically in the chiral limit
for $\xi=4\, (5)$ we find $\Delta m_\pi\simeq 3.6\, (4)$  MeV
to be compared with the experimental value
$\Delta m_\pi\simeq 4.6$  MeV.

The  coefficients $L_i$ and $C$ 
have been previously calculated    using  different  approximations.
For example, in Refs.~\cite{Ecker:1988te,Peris:1998nj} 
these coefficients were calculated from an effective 
 theory of resonances,
 showing a good agreement with the experimental data.
It would be interesting to study the relation between
 the approach presented
here with  those of Refs.~\cite{Ecker:1988te,Peris:1998nj}.

\section{Conclusions}

\begin{figure}[htbp]
  \centering
  \psfrag{c6}{\small{$c_6$}}
  \psfrag{mA1}{\small{$M_{A_1}$}}
  \psfrag{Fpi}{\small{$F_{\pi}$}}
  \psfrag{Fr}{\small{$F_{\rho}$}}
  \psfrag{g1}{\small{$g_{\rho\pi\pi}$}}
  \psfrag{dmPi}{\small{$\Delta m_{\pi}$}}
  \psfrag{x}{$\xi$}
  \psfrag{te}{Th/Exp}
  \includegraphics[width=10cm]{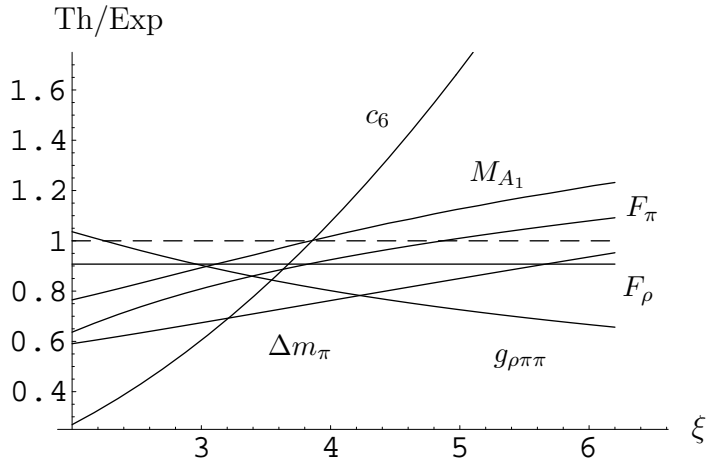}
  \caption{\textit{Predictions of the  model
for  some physical quantities   as a function of $\xi$
divided by their experimental value.  We have taken $M_q=0$.}}
  \label{Figresult1}
\end{figure}

We have presented a 5D model that describes some of the properties
of QCD related to chiral symmetry breaking. Alike large-$N_c$ QCD,
this model is defined by a set of  infinite weakly coupled
resonances. The model depends only on  one  parameter, $\xi$,
related to the quark condensate (apart   from the other 3
parameters of the model that are fixed by the  3 parameters that
define QCD: the mass gap $\Lambda_{\rm QCD}$, $M_q$, and $N_c$).
We have obtained predictions for the masses and decay constants of
the vector, axial-vector and PGB mesons. These predictions  are in
good agreement with the experimental data. A summary of some of
the results is given in Table~1 and Fig.~\ref{Figresult1} that
shows that, within a 30$\%$, they agree with the data.

The 5D gauge invariance of the model leads to  interesting 
sum rules among the
couplings and masses of the resonances from which we obtain 
$M^2_\rho \simeq  3\,
g_{\rho\pi\pi}^2 F_\pi^2$, $F_\rho\simeq \sqrt{3}\, F_\pi$, and the
vanishing of the BR of $a_1$ into $\pi\gamma$ at the
tree-level. Another prediction of the model is the realization
of VMD in the electromagnetic form factor of the pion.

Since the results presented here depend on the AdS$_5$ metric Eq.~(\ref{ads}),
one can wonder whether the results are robust under possible
deviations from AdS. For example, if the backreaction on the
metric due to $\langle\Phi\rangle^2$  or other possible
condensates are included  in the model, we expect the warp factor
$a(z)$ to depart from AdS in the IR. Nevertheless if we want the
theory to be almost  conformal in the UV, the warp factor 
for $z\ll L_1$  (where  $1/L_1$ gives the mass gap) must  behave  as
\begin{equation}
a(z)\simeq \frac{L}{z} \left[1+\sum_i
c_i\left(\frac{z}{L_1}\right)^{d_i}\right]\, , \label{restriction}
\end{equation}
where $c_i$ are numerical constants related to the singlet condensates
$\langle{\cal O}_i\rangle$ and $d_i={\rm Dim}[{\cal O}_i]$. In QCD
$d_i\geq 4$. Eq.~(\ref{restriction}) implies that only for values
of $z$  quite close to $L_1$ the metric will deviate from AdS. Therefore,
unless the coefficients $c_i$ are very large, we  do not expect
large deviations from our results. The  $c_i$, however, are
restricted by  the curvature of the space ${\cal R}$. We have
checked that for  ${\cal R}\sim 1/L_1^2$, our results are not
substantially modified by deformations of the AdS metric in the IR. 
As an example,  we have compared some of
our results to those with the  metric of 
Ref.~\cite{Karch:2002sh},
\footnote{This is the induced metric on the D7-brane 
on which the gauge bosons propagate in  Ref.~\cite{Karch:2002sh}.}
 $a(z)=\frac{\pi L}{2L_1 \sin[{\pi z}/{2L_1}]}$,
and we have found that the differences 
are smaller than $10\%$.
We can then conclude that  more realistic 
 string constructions of QCD,
such as those of Refs.~\cite{Kruczenski:2003be}, 
must lead to quantitatively similar results.

\vskip2cm

{\bf Note Added:}  While writing this paper, it appeared
Ref.~\cite{Erlich:2005qh} that proposes the same  5D
 model to study the  properties of the QCD hadrons.

\section*{Acknowledgements}
\label{sec:acknowledge}

We would like to thank Santi Peris for very useful
discussions. This work  was  partly supported   by the MCyT and
FEDER Research Project FPA2002-00748 and DURSI Research Project
2001-SGR-00188.
The work of LD was supported 
by the Spanish Education Office (MECD) under 
an FPU scholarship.

\section*{Appendix. 
Chiral symmetry breaking induced by  an operator of dimension $d$}
\label{ngral}

In this appendix we give  the expression for $\Pi_A$
in the different limits studied in the text for the  case
in which the breaking of the chiral symmetry arises
from a VEV of a  scalar $\Phi$ with an arbitrary 5D mass $M_\Phi$.
This
corresponds in the CFT
to turning on an operator of dimension $d=\sqrt{4+M^2_\Phi L^2}+2$.

For small momentum we have
$\Pi_A(p)=\Pi_A^{(0)}(0)+M_qL_1\Pi_A^{(1)}(0)+p^2\Pi^\prime_A(0)+\cdots$ where
in the limit $\xi \gg 1$:
\begin{eqnarray}
\Pi_A^{(0)}(0)&\simeq&\frac{2^{(1-1/d)} d^{(1-2/d)}\pi}
{\sin(\pi/d)\,\Gamma(\frac{1}{d})^2}\frac{\tilde
  N_c \,\xi^{2/d}}{L_1^2}\, ,\\
\Pi_A^{(1)}(0)&\simeq&
\frac{2^{(1-1/d)}\Gamma(\frac{2+d}{2d})\Gamma(\frac{3}{d})}{d^{(1-2/d)}\Gamma(\frac{4+d}{2d})}
\frac{\tilde N_c\,   \xi^{1-2/d}}{L_1^2}
\left (1-\frac{d^{2/d}\Gamma(\frac{4+d}{2d})\Gamma(\frac{4}{d})}
{2^{1/d}\Gamma(\frac{6+d}{2d})\Gamma(\frac{1}{d})} \frac{1}{\xi^{2/d}}\right)
\, ,\\
\Pi'_A(0)&\simeq& -\tilde N_c \ln\frac{L_0}{L_1}
-\tilde N_c\left[\ln \xi^{1/d}
+\frac{\gamma+
\psi\left(\frac{2+d}{2d}\right)-\psi\left(\frac{2}{d}\right)
-\psi\left(\frac{1}{d}\right)-\ln\frac{d^2}{2}}{2 d}\right]\, ,
\end{eqnarray}
 where $\psi(x)=\Gamma'(x)/\Gamma(x)$.

In the large momentum limit we have
\begin{equation}
\Pi_A(p^2)=-p^2\left[
\frac{\tilde N_c}{2} \ln(p^2L_0^2)+\frac{c_{2d}}{p^{2d}}+\cdots\right]\, ,
\end{equation}
where
\begin{equation}
c_{2d} =
-\frac{d\, \sqrt{\pi}}{2(d-1)}\frac{\Gamma(d)^3}{\Gamma(d+\frac{1}{2})}
\frac{\tilde N_c\, \xi^2}{L_1^{2d}}\, .
\end{equation}

From the above expressions we can derive
$L_5$ and $L_{10}$:
\begin{equation}
L_5\simeq \tilde N_c
\frac{2^{(-2-2/d)}\pi\Gamma(\frac{2+d}{2d})\Gamma(\frac{3}{d})}
{\sin(\pi/d)\,\Gamma(\frac{4+d}{2d})\Gamma(\frac{1}{d})^2}
\left (1-\frac{d^{2/d}\Gamma(\frac{4+d}{2d})\Gamma(\frac{4}{d})}
{2^{1/d}\Gamma(\frac{6+d}{2d})\Gamma(\frac{1}{d})} \frac{1}{\xi^{2/d}}\right)
\; ,
\end{equation}
\begin{equation}
L_{10}\simeq
-\frac{\tilde N_c}{4}\left[\ln \xi^{1/d}
+\frac{\gamma+
\psi\left(\frac{2+d}{2d}\right)-\psi\left(\frac{2}{d}\right)
-\psi\left(\frac{1}{d}\right)-\ln\frac{d^2}{2}}{2 d}\right]\, .
\label{L10d}
\end{equation}
The parameter $L_{10}$  allows to calculate the Peskin-Takeuchi
$S$ parameter \cite{Peskin:1991sw} of Technicolor-like theories
where $F_\pi=246$ GeV triggers the breaking of the electroweak
symmetry. This is  given by $S=-16\pi L_{10}$. Electroweak
precision tests tell us that $S\lappeq 0.3$, a constraint difficult
to be satisfied in the present models
\cite{Csaki:2003zu,Barbieri:2003pr}. From Eq.~(\ref{L10d}) we can
derive the dependence of $S$ on $d$. For a fixed value of $F_\pi$
we find that the dependence on $d$ is very weak, and $S$ changes
only a few per cent when varying $d$. This implies that $S$ in
these type of models is always  sizable.


\end{document}